\documentclass[aps,prl,preprint,groupedaddress]{revtex4-1}
\usepackage{graphics}  % needed for figures
\usepackage{amsmath}   % for math
\usepackage{amssymb}   % for math

\begin{document}

\title{Halbach arrays at the nanoscale from chiral spin textures}

\author{Miguel~A.~Marioni}
\email[Corresp.~author.~E-mail:~]{miguel.marioni@empa.ch}
\affiliation{Empa, Swiss Federal Laboratories for Materials Science and Technology, CH-8600 D\"{u}bendorf, Switzerland.}
\author{Marcos Penedo}
\affiliation{Empa, Swiss Federal Laboratories for Materials Science and Technology, CH-8600 D\"{u}bendorf, Switzerland.}
\author{Mirko~Ba\'{c}ani}
\affiliation{Empa, Swiss Federal Laboratories for Materials Science and Technology, CH-8600 D\"{u}bendorf, Switzerland.}
\author{Johannes~Schwenk}
\altaffiliation{Center for Nanoscale Science and Technology, National Institute of Standards and Technology, Gaithersburg, MD 20899, USA.}
\affiliation{Empa, Swiss Federal Laboratories for Materials Science and Technology, CH-8600 D\"{u}bendorf, Switzerland.}
\author{Hans~J.~Hug}
\altaffiliation{Department of Physics, University of Basel, CH-4056 Basel, Switzerland.}
\affiliation{Empa, Swiss Federal Laboratories for Materials Science and Technology, CH-8600 D\"{u}bendorf, Switzerland.}

\date{\today}

\begin{abstract}
Mallinson's idea\cite{1Mallinson:1973} that some spin textures in planar magnetic structures could produce an enhancement of the magnetic flux on one side of the plane at the expense of the other gave rise to permanent magnet configurations known as Halbach\cite{2Halbach:1980} magnet arrays.
Applications range from wiggler magnets in particle accelerators and free electron lasers\cite{3Balal:2015}, to motors\cite{4Zhu:2017}, to magnetic levitation trains\cite{5Post:2000}, but exploiting Halbach arrays in micro- or nanoscale spintronics devices requires solving the problem of fabrication and field metrology below 100\,$\mu$m size\cite{6Taylor:2008}.
In this work we show that a Halbach configuration of moments can be obtained over areas as small as $1\times 1\,\mu$m$^2$ in sputtered thin films with N\'{e}el-type domain walls of unique domain wall chirality, and we measure their stray field at a controlled probe-sample distance of $12.0\pm 0.5$\,nm.
Because here chirality is determined by the interfacial Dyzaloshinkii-Moriya interaction\cite{7MoreauLuchaire:2016} the field attenuation and amplification is an intrinsic property of this film, allowing for flexibility of design based on an appropriate definition of magnetic domains.
100\,nm-wide skyrmions illustrate the smallest kind of such structures, for which our measurement of stray magnetic fields and mapping of the spin structure shows they funnel the field toward one specific side of the film given by the sign of the Dyzaloshinkii-Moriya interaction parameter $D$.
\end{abstract}

% pacs{}

\maketitle

Figure 1a illustrates the remarkable property of Halbach magnet arrays of channeling the magnetic flux such that it largely emerges out of a defined side of the structure (the top side, in this instance).
Such characteristics provide devices with a strong static magnetic field at room temperature.
The arrangement of sources of magnetic field in the Halbach array and in thin-film spin textures comprising through-thickness perpendicular-magnetization domains with N\'{e}el-type walls (Figure 1b) has similar characteristics.
Analogous to the former case, magnetic charges through the film thickness flank the domains with up/down magnetization, in this case arising from a non-vanishing magnetization divergence in the N\'{e}el wall.
Importantly, though the domains' surface charges on top and bottom sides of the structure differ in sign, the volume charges of a given wall have the same sign through-thickness, leading to the asymmetric disposition of field sources necessary for the field enhancement or attenuation.
A cursory inspection of the arrows indicating the direction of rotation of the magnetic moments across a wall in Figure 1b reveals the connection between a well-defined chirality of the N\'{e}el wall magnetization and the topology of magnetic charges characterizing a Halbach array.
Neither N\'{e}el walls nor walls with unique chirality throughout the film are a priori given in thin films with perpendicular magnetization, but a sufficiently strong Dzyaloshinskii-Moriya (DM) interaction in thin film interfaces\cite{7MoreauLuchaire:2016} can stabilize both, determining the preponderance of N\'{e}el walls over Bloch walls, and removing the degeneracy of their chirality\cite{7MoreauLuchaire:2016,8Chen:2013,9Boulle:2016,10Chen:2015}.
Notable examples include Ir/Ni\cite{7MoreauLuchaire:2016,8Chen:2013}, Ir/Co\cite{7MoreauLuchaire:2016,10Chen:2015}, Pt/Ni\cite{8Chen:2013}, Pt/Co\cite{7MoreauLuchaire:2016,10Chen:2015}, Mn/W\cite{11Ferriani:2008}, Ir/Fe and Ir/(FePd)\cite{12Romming:2013}, and Ta/Co/TaOx\cite{13Jiang:2015}.
We focus on an Ir/Co/Pt system\cite{7MoreauLuchaire:2016} and measure the stray fields of different domain configurations to show that it is possible to obtain a local enhancement or attenuation of the magnetic field at the scale of few 100\,nm, and control it with the sign of the DM interaction.
\begin{figure}
  \includegraphics{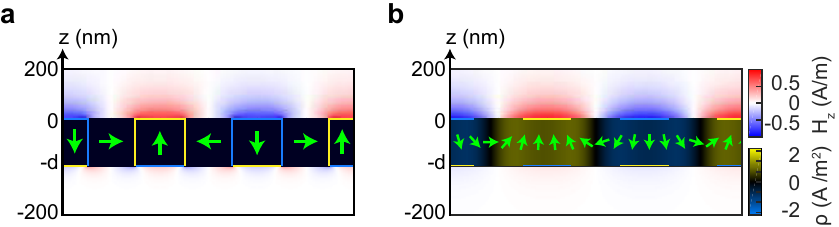}
\caption{{\bf Field enhancement in Halbach arrays.}
{\bf a}. Halbach array showing larger fields above- and smaller fields below the magnet array consisting of 200\,nm square magnets with a magnetization of 1\,A/m (center band, black).
The blue and yellow lines in the center band indicate surface magnetic charges of negative and positive sign, respectively.
{\bf b}. Perpendicular domain pattern in a film of 200\,nm thickness with chiral N\'{e}el walls with a wall width of 150\,nm.
The unique chirality for the N\'{e}el walls is necessary to establish spin textures with the same topology as {\bf a}.
}
  \label{fig1}
\end{figure}

To quantify the stray fields of the Halbach structures in application-relevant conditions (industry-standard sputtered, oxidation-protected thin film systems) is challenging at high resolution, of the order of 10\,nm.
High resolution magnetic force microscopy (MFM) turns out to be convenient for various reasons and we use it here.
Briefly, with the appropriate tip calibration MFM is quantitative\cite{14Vanschendel:2000}, and can measure all stray field components emanating from a magnetic sample.
This is possible also when the sample is an oxidation-capped thin film on a substrate, for a wide range of stray fields, and in external fields applied to select the desired micromagnetic state.
Importantly also, the distance between probe and sample can be measured and controlled\cite{15Schwenk:2014,16Schwenk:2015,17Zhao:2017} to within 1\,nm, an ability without which highly precise magnetic field measurements could not be ascribed to a specific source with useful accuracy.
Nevertheless the same observations from this work could be made with other quantitative field-measurement methods, notably NV-diamond tip magnetometry\cite{6Taylor:2008}, at least as long as the total fields remain small and provided the sample-NV center distance can be ascertained  and controlled.

Figure 2a is a MFM measurement of spontaneous domains in an [Ir/Co(0.6nm)/Pt]$\times 5$ film\cite{7MoreauLuchaire:2016} at $12.0\pm 0.5$\,nm above the sample surface.
The stray fields from up and down magnetized domains in the film (Figure 2d) are sensed as a shift in the cantilever resonance frequency (indicated by the colors in Figure 2a).
As expected from the thickness loss\cite{0Buschow_1993,18Bertram:1994,0Porthun:1998} the contrast is largest across the domain walls (narrow boundaries with $\delta f=0$ , in green), and the signal decreases in amplitude away from a wall\cite{18Bertram:1994,19Joshi:2011}.
The interaction between the stray fields and the MFM tip (Figure 2b top panel) depends of the a priori unknown distribution of magnetic moments of the tip.
This distribution is measured in the calibration of the tip\cite{14Vanschendel:2000}, and is an essential part of the tip-calibration function.
With appropriate account of geometrical factors and space propagation, it can be visualized as the tip's stray field amplitude on a reference plane parallel to the scan plane, as in Figure 2c, for instance.
In general, knowledge of the calibration function allows calculating the frequency shift arising from a given stray field in the specific calibrated system.
For Figure 2e-g we use the calibration function in conjunction with stray fields calculated from the domain pattern depicted in Figure 2d, which underlays the measurement in Figure 2a, to see how different domain wall characteristics would affect the measurement.
Thus, for Figure 2e we calculate the contrast expected for the magnetization pattern in Figure 2d when the domain pattern's walls are N\'{e}el type with clockwise (cw) chirality, and characterized by exchange stiffness and measured values of anisotropy and saturation magnetization\cite{20Rohart:2013}.
Similarly, Figures 2f, g display the simulated measurement when the walls are respectively of Bloch type, or N\'{e}el type with counter clockwise (ccw) chirality.
Note that the latter corresponds to the MFM contrast that would be obtained on the substrate side of our sample.
\begin{figure}[h]
  \includegraphics{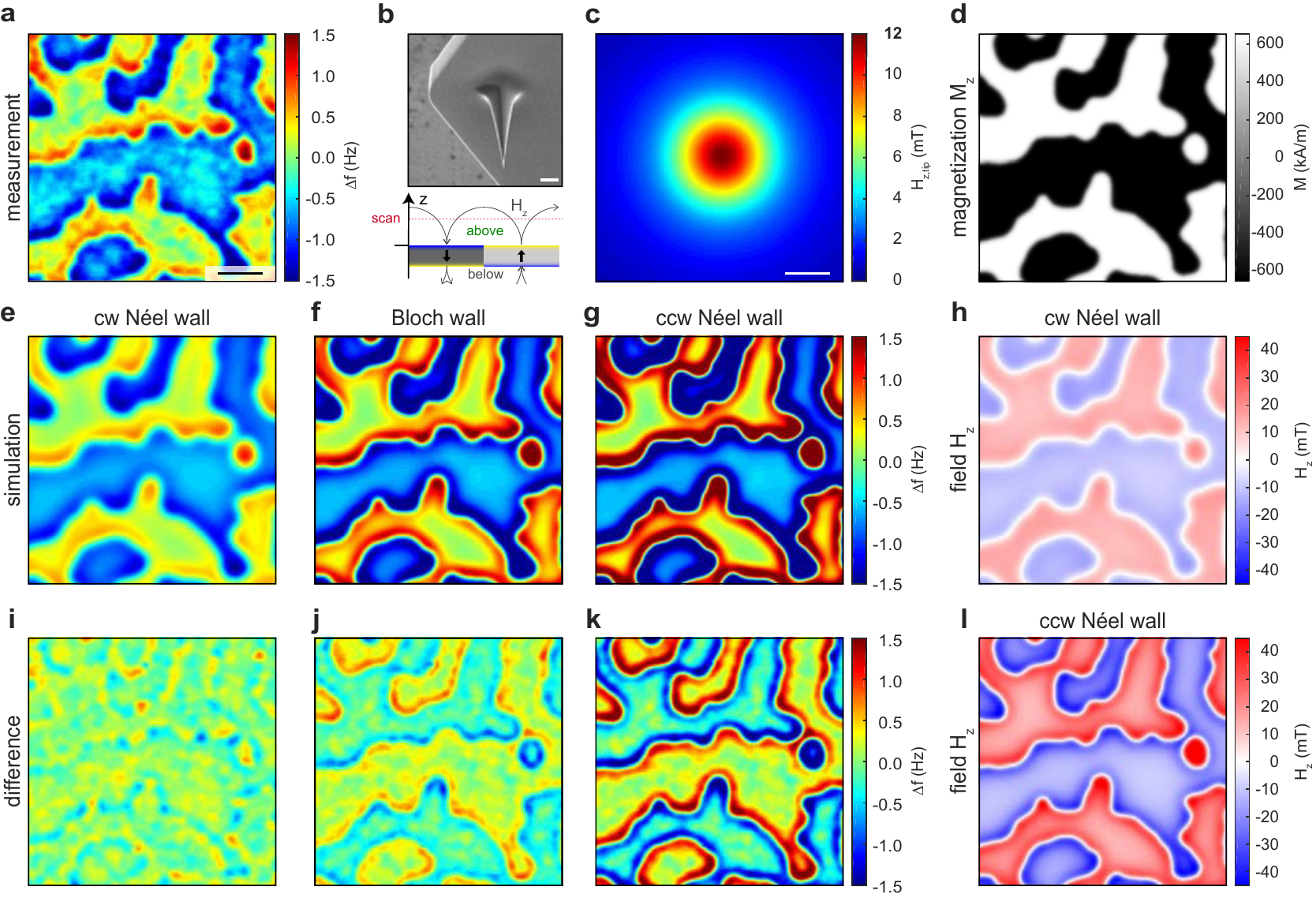}
\caption{
{\bf Magnetic domains in [Ir/Co(0.6nm)/Pt]$\times 5$}.
{\bf a} Measured MFM signal.
Scale bar 200\,nm.
{\bf b} He-microscope image of the cantilever tip, which scans $12.0\pm 0.5$\,nm over the sample indicated by the inset below the image of the tip.
Scale bar 2$\mu$m.
{\bf c} Equivalent tip magnetic field (a measure of the tip magnetic charge distribution) on a scan plane 12\,nm away from the tip apex.
Scale bar 20\,nm.
{\bf d} Up/down domain pattern underlying a.
{\bf e} Simulation of and MFM measurement for a pattern of perpendicular domains as in a and clockwise N\'{e}el walls.
{\bf f} Analogous simulation using Bloch walls.
{\bf g} Analogous simulation using counter clockwise N\'{e}el walls.
{\bf i} Difference between measurement {\bf a} and simulation {\bf e}.
{\bf j}, {\bf k} Corresponding differences between {\bf a} and {\bf f}, or {\bf a} and {\bf g} respectively.
{\bf h}, {\bf l} Magnetic field corresponding to e and g, respectively.
Map {\bf h} depicts the $z$-field on the measured plane, and map {\bf l} depicts the opposite side of the film.
Panel {\bf d}-{\bf l}  dimensions same as {\bf a}.
}
  \label{fig2}
\end{figure}

Comparison of measurement and simulation, in Figures 2i-k for each case, respectively, indicates that the system possesses a positive value of the DM coefficient $D$ (clockwise magnetization rotation) and consequently, that the field on the measurement plane (Figure 2h, above the sample per Figure 2b) is attenuated compared with the field on the symmetric point on the other side of the film (Figure 2l, underneath the film per Figure 2b).
Note that the maximum field amplitude 12\,nm above the film is approximately 37\% of the corresponding value at 12\,nm below the film.
At domain sizes as narrow as 200\,nm and a wall width of 33\,nm, therefore, we see that Halbach-like magnetic structures can be created.

Provided their stability was warranted, smaller sized structures featuring Halbach-like attenuation/enhancement of the stray field could find application in devices of dimensions ranging down to few tens of\,nm\cite{21Yu:2012}.
A prototypical structure in this case is the magnetic skyrmion\cite{22Bogdanov:1994}.
On account of having the same topology as a chiral bubble domain\cite{23Nagaosa:2013} we can expect different  field amplitudes on either side of the film also in this case.
In first instance the difference would be revealed in the frequency shift ($\delta f$) that our MFM registered for such objects.
To quantify it we simulate skyrmions in our multilayer and utilize the tip calibration function to obtain corresponding frequency shift ($\delta f$) patterns for our specific MFM.

The spin profiles from skyrmions (Figure 3a) that are stable in our multilayer for various values of D result in different MFM contrast ($\delta f$) profiles (Figure 3c. Left side panel: half-profile for $D>0$, right panel: half-profile for $D<0$).
Because the skyrmions for larger $|D|$ values are wider, the maximum MFM contrast generated by them is stronger (Figure 3b).
Moreover, skyrmions with negative $D$, characterized by a counter clockwise rotation of their magnetization direction generate a considerably larger maximum contrast.
\begin{figure}
  \includegraphics{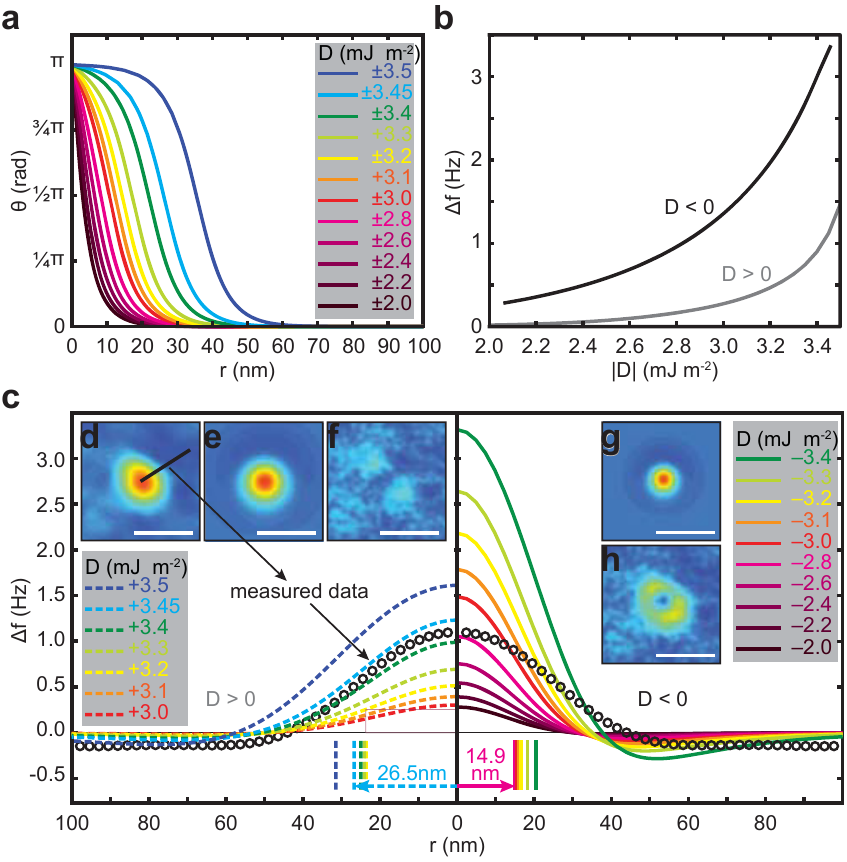}
\caption{
{\bf Skyrmion profiles in [Ir/Co(0.6nm)/Pt]$\times 5$}.
{\bf a} Calculated magnetization orientation (half-) profiles.
{\bf b} Skyrmion peak MFM contrast as function of the DM interaction parameter $D$.
Negative $D$-values result in a larger center contrast (black line) than positive $D$-values (grey line).
{\bf c} Simulated MFM frequency shift profiles of skyrmions with positive- (left, dotted lines) and negative $D$ (right, solid lines).
The half-width/half-maximum values are indicated by the vertical segments below the abscissa.
The black circles depict an experimental cross section.
{\bf d} Measured skyrmion.
{\bf e} Simulated skyrmion for $D>0$ chosen to match the experimental peak contrast as per {\bf b}.
{\bf f} Point-wise difference between {\bf d} and {\bf e}.
{\bf g} and {\bf h} Same as {\bf e} and {\bf f}  for $D<0$ chosen to match the experimental peak contrast as per {\bf b}.
Scale bars {\bf d}-{\bf h}: 100\,nm; same color scale.
}
  \label{fig3}
\end{figure}

From Figure 3b, the peak contrast of the measured skyrmion can be expected for skyrmions of two different values of $D$, which opposite sign.
However, only for one of those values the skyrmion widths are matched as well, and this condition uniquely determines the sign of $D$.
The difference between skyrmions of either sign of $D$ becomes apparent when comparing the experiment results (Inset Figure 3d) with the modelled skyrmion contrast for positive $D$ (inset Figure 3e), and negative $D$ (inset Figure 3g).
Experiment and simulation are matched for $D>0$ (Inset Figure 3f and h show the differences between the experimental results Figure 3d and the simulation for positive and negative $D$, respectively).
Measured cross-sectional data (black circles in Figure 3c) displayed with the simulated MFM profiles for positive $D$ (left side) and negative $D$ (right side) show that for the latter is not possible to simultaneously match the peak contrast and extension.
Thus the profile for $D=+3.42$\,mJ/m$^2$ with a half-width/half-maximum-radius of 26.5\,nm matches the measured skyrmion profile data well, but for D=-2.80 mJ/m2 the peak contrast is matched but not the half-width half-maximum (HWHM) radius.

The results allow us to paint a complete picture of the skyrmion stray fields.
Figures 4b-e display the measured stray field vector components of the skyrmion displayed in Figure 4a (and inset Figure 3d).
Figure 4f displays the clockwise hedgehog orientation of the skyrmion spins for $D=+3.42$\,mJ/m$^2$ resulting in the best fit of the modeled to the measured skyrmion MFM data and correspondingly the modeled field (Figure 4e) to that deconvolved from the MFM data (Figure 4b).
Conversely, Figure 4h displays the field over the skyrmion on the opposing side of the film, or for a counter clockwise hedgehog orientation of the skyrmion spins resulting in a stray field that is much stronger than the measured one.
A weaker stray field equal to that obtained from the deconvolution of the measured data is obtained if a smaller negative $D=-2.80$\,mJ/m$^2$ is used.
The corresponding hedgehog spin orientation is shown in Figure 4i.
This however results in a modeled MFM skyrmion image with a HWHM-radius of 14.9\,nm considerably smaller than that observed (Figure 3c-h).
The picture that emerges is one of a pointed source of stray field at negative $z$-positions that is attenuated at the corresponding positive $z$, as anticipated from considering Mallison and Halbach's arrays.
This high-resolution characterization of multilayers with strong interfacial Dzyaloshinskii-Moriya interaction at scales of few 10\,nm, and of predominantly one-sided localized sources of magnetic field shows a way to implement Halbach magnet arrays at the nanometer scale.
Because an amplified field produced on one side of the structure would be attenuated on the other, and it could be integrated into other thin film structures, it constitutes a tool for densely packing fields for spintronics devices.
\begin{figure}
  \includegraphics{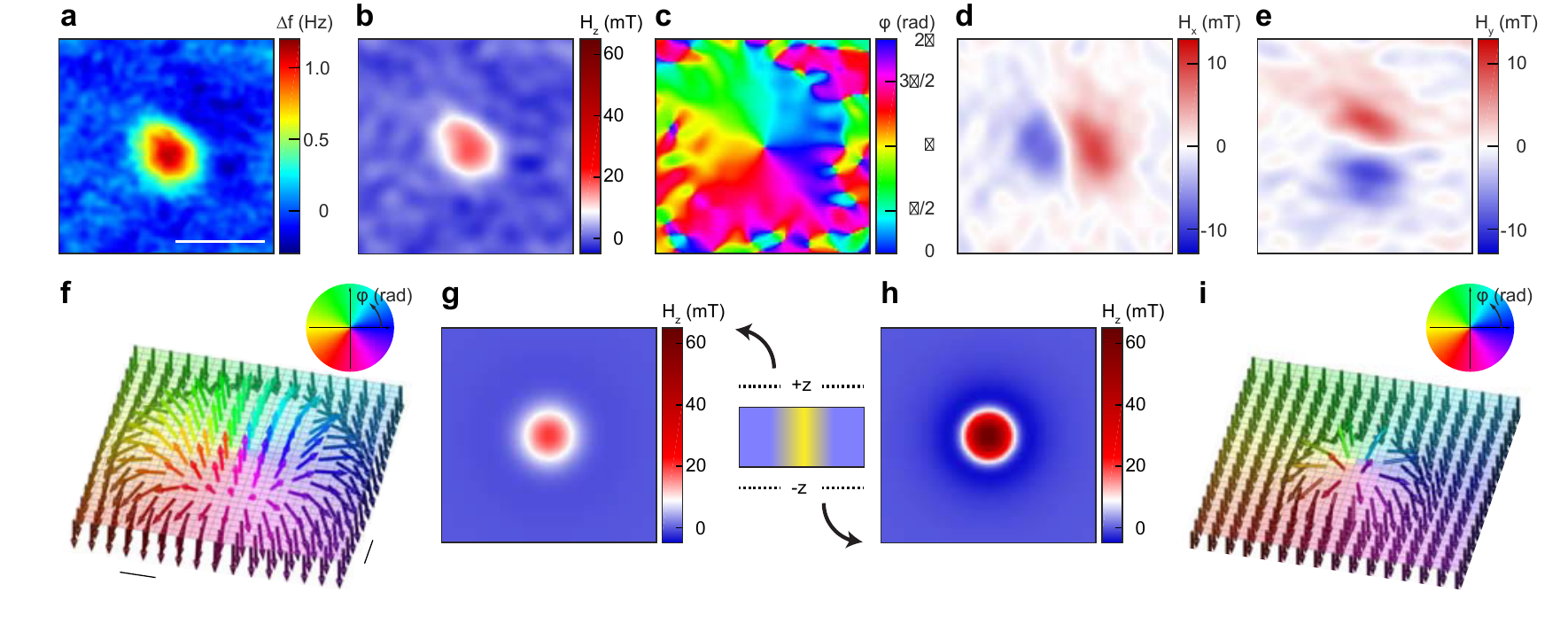}
\caption{
{\bf Skyrmion profiles in [Ir/Co(0.6nm)/    Pt]$\times 5$}.
{\bf a} MFM measurement of a skyrmion stray field $12.0\pm 0.5$\,nm above the surface.
Scale bar 100\,nm.
{\bf b}-{\bf e} Stray magnetic field components from the deconvolution of a using the tip calibration function, respectively: $z$-component, azimuthal angle $\phi$, $x$-component, and $y$-component.
{\bf f} Spin texture of the skyrmion measured in {\bf a}. Scale bars: 10\,nm.
{\bf g} Quantitative simulation of the $z$-component of the stray field at $z =+12.0$\,nm using $D=+3.42$\,mJ/m$^2$.
{\bf h} Same as {\bf g} at $z=-12.0$\,nm for $D=-3.42$\,mJ/m$^2$.
Panel dimensions {\bf b}-{\bf e}, {\bf g}, {\bf h} same as {\bf a}.
{\bf i} Same as {\bf f}  fitting b with opposite chirality: $D=-2.80$\,mJ/m$^2$.
Same panel dimensions as {\bf f}.
}
  \label{fig4}
\end{figure}

\vspace{1cm}\noindent{\bf Film preparation and magnetometry.}
We fabricated the samples by magnetron sputtering on naturally oxidized Si(100), depositing at room temperature.
The systems' structure is Pt(10nm)/Co(0.6nm)/Pt(1nm)[Ir(1nm)/Co(0.6nm)/Pt(1nm)]$\times$5/Pt(3nm).
By vibrating sample magnetometry (VSM) at a Quantum Design PPMS we obtain the anisotropy $K_{\rm u} = 414$\,kJ/m$^3$, and the Co layers' magnetization of 653.6\,kA/m.
The effective anisotropy is thus $K_{\rm eff} = 146$\,kJ/m$^3$.

\vspace{1cm}\noindent{\bf MFM.}
For the microscopic characterization we used a room temperature magnetic force microscope operated in vacuum ($10^{-6}$\,mbar) using the first flexural mode of the cantilever.
The probes are 0.7\,N/m type-SS-ISC cantilevers (tip radius smaller than 5\,nm) from team Nanotec GmbH coated in-house.
We measure at $12.0\pm 0.5$\,nm from the sample surface by actively controlling the tip sample distance capacitively using the second cantilever flexural oscillation mode\cite{16Schwenk:2015}.
For the tip calibration with which to render the measurements quantitative we employ the method outlined by van Schendel et al.\cite{14Vanschendel:2000} which relies on known patterns of magnetization.
We utilize domains from Pt(10nm)[Co(0.6)nm/Pt(1nm)]$\times 5$/Pt(3nm) obtained after demagnetization with an oscillatory in-plane field.
The tip calibration function is optimized for 2601 measured domain patterns (see supporting material).

\vspace{1cm}\noindent{\bf Simulation.}
For the data processing and imaging we use the qMFM Matlab analysis package (\!\!\texttt{ http://qmfm.empa.ch}).
For modelling the skyrmions we numerically solved the Euler equation for the energy functional from Bogdanov and Hubert\cite{22Bogdanov:1994} modified to account for our multilayer stack geometry (Supplemental Material).

\vspace{1cm}\noindent{\bf Acknowledgements}
This work was supported by Empa through the internal grant ``Skyrmions''.
Additional support was provided by SNF for J.S.~through Sinergia `Understanding nanofriction and dissipation across phase transitions' and for M.B. through Sinergia `Complex Oxides'.
We gratefully acknowledge CCMX for support through the ``Quantitative Magnetic Force Microscopy platform (qMFM)'', in which the Matlab analysis platform (\texttt{ http://qmfm.empa.ch}) was programmed by Zoe Goey, Sven Hirsch and Gabor Sz\'{e}kely.
We also would like to thank Sara Romer and Alexandre Guiller for contributions to the film fabrication and magnetometry.
We acknowledge Sasa Vranjkovic for the design and construction of instrument components, and Fabio La Mattina for He FIB characterization.

\end{document}